# Kinetics of transfer of volatile amphiphiles (fragrances) from vapors to aqueous drops and vice versa: Interplay of diffusion and barrier mechanisms

Krassimir D. Danov,[*] Theodor D. Gurkov, Rumyana D. Stanimirova, Ralitsa I. Uzunova

*Department of Chemical & Pharmaceutical Engineering, Faculty of Chemistry & Pharmacy, Sofia University, 1164 Sofia, Bulgaria*

* Corresponding author.
  E-mail address: KD@LCPE.Uni-Sofia.BG (K.D. Danov)

ABSTRACT

Subject of this work is to investigate the kinetics of mass transfer of volatile amphiphiles from their vapors to aqueous drops, and from the saturated aqueous drop solutions to air. The used amphiphiles are benzyl acetate, linalool, and citronellol, all of which have low saturated vapor pressures, appreciable solubility in water, and well pronounced surface activity. The adequate theoretical processing of the equilibrium surface tension, $\sigma$, isotherms is applied to construct the two-dimensional equation of state, which relates $\sigma$ to the adsorption, $\Gamma$, at the interface. The measured surface tension relaxations with time $t$ in the regimes of adsorption from vapor and evaporation from drop combined with the equations of state provide quantitative information on the change of adsorption because of the volatile amphiphile mass transfer across the surface. The theoretical analysis of the diffusion and barrier mechanisms in the case of adsorption from vapor to the aqueous drop shows that the mixed barrier-diffusion control in the vapor and diffusion control in the drop describe experimental data. The obtained values of the adsorption rate constants are six orders of magnitude larger than those for hexane and cyclohexane reported in the literature. The regime of evaporation from aqueous amphiphile solution drop follows the convection-enhanced adsorption mechanism with desorption rate constant in the vapor affected by the simultaneous water evaporation and amphiphile desorption. The water evaporation suppresses the evaporation of linalool and accelerates desorption of benzyl acetate and citronellol. From viewpoint of applications, the obtained physicochemical parameters of the studied three fragrances can help for better understanding of their performance in shampoo systems and perfumes. From theoretical viewpoint, the result show that by introducing an effective amphiphile desorption rate constant it is possible to quantify the complex volatile amphiphile desorption accompanied with the water evaporation.

1. **Introduction**

The evaporation and condensation of volatile components have wide industrial applications including inkjet printing [1,2], bio-sensing and thermal electronic devices [3–6], spray cooling and coating [7–10], etc. More than 26.3 billion dollars industry [11,12] produces the familiar fragrances and flavors that surround us in everyday life. The volatile organic compound analysis is widely used in medicine for the disease detection and therapeutic monitoring [13,14]. Fragrances and malodors are ubiquitous in the environment and their detection has a broad range of civilian, military, and national security applications [15]. Thousands of flavor and fragrance compounds have been characterized by two-dimensional gas chromatography combined with mass spectroscopy [11].

The volatile amphiphiles pertain to a class of volatile organic compounds which have low solubility in water, well solubility in alcohols, ethers, and partial solubility in some oils, and a wide range of vapor pressure at room temperature (up to thousands of Pa). They adsorb at the water/vapor and water/oil interfaces, reduce the interfacial tension and change the interfacial rheology. Even at very low concentrations, they are used in the shampoos, lotions, and detergents, the perfumes can change the size of micelles and act on the bulk viscosity of shampoos as thinning or thickening agents [16,17]. Except of the well-studied physicochemical properties of volatile amphiphiles (solubility, vapor pressure, etc.), for the purpose of predictive modeling of their interfacial properties in complex mixtures also the following information is needed for the individual volatile amphiphiles: first, the adsorption isotherms at vapor/solution and oil/solution interfaces give information for the equilibrium properties of adsorbed molecules. The dynamics of adsorption and desorption of volatile amphiphiles from the interfaces to the surrounding media defines the characteristic relaxation times. The understanding of the mechanisms of adsorption (diffusion, barrier, convective-enhanced, or mixed) help to obtain the possible ways for the control of the emulsification and foam properties of the complex mixed solutions of practical interest. Unfortunately, this information is difficult to be found in the literature or missing at all.

The surface tension isotherms of aqueous solutions of 10 monoterpene alcohols at 20 °C are measured in Ref. [18]. The authors processed experimental data using various adsorption models of localized and non-localized adsorption and obtained the respective physicochemical parameters. The qualitative pictures of alkane adsorption at the interface have been outlined using molecular dynamics simulations [19,20]. The dynamics of adsorption of short-chain alkanes (pentane, hexane, heptane, octane) from the saturated vapor to the water drop surface [21,22] shows that at initial times (up to 15 minutes) these volatile amphiphiles form monolayers at the interface. With the increase of time, the interfacial tension gradually decreases and the multilayer adsorption is detected. Subsequently, the molecular adsorption transfers into a condensation leading to a thin alkane film at the drop surface. The respective adsorption processes from vapor are barrier-diffusion controlled and the adsorption rate constants depend considerably on temperature.

The effect of the co-adsorption of hexane from the vapor phase at the surface of aqueous drop with dissolved nonionic and ionic surfactants and proteins is studied in Refs. [23–31]. The co-adsorption of hexane is most pronounced for surfactant concentrations below the critical micelle concentration of surfactant and the multilayer adsorption is again observed at long times. The effect of fluorocarbon vapors on the adsorption dynamics of phospholipid monolayers at the aqueous drop is studied in Refs. [32,33]. The authors showed that the mechanism of alkane adsorption from vapor is barrier controlled and obtained the adsorption (condensation) and the desorption (evaporation) rate constants.

In the present study, we investigate the dynamics of adsorption of benzyl acetate, linalool, and citronellol from vapors to the aqueous drops and the subsequent desorption from their saturated aqueous drop solutions to the air. The three volatile amphiphiles have considerably lower saturated vapor pressure (from 7 to 22 Pa) compared to that of pentane (> 68 kPa) quoted in the literature [22], and have limited (not negligible) solubilities in water. The measured equilibrium adsorption and surface tension isotherms (Section 3) help us to relate the dynamic surface tension with the adsorption at the given moment. In the regimes of adsorption from saturated vapors to the aqueous drop surface, the mechanism of adsorption corresponds to the mixed barrier-diffusion control, which allows obtaining the adsorption and desorption rate constants (Section 4). In the regime of evaporation, the drop is in contact with the ambient atmosphere (where vapours of volatile amphiphiles are absent). The simultaneous water and amphiphile evaporation changes the desorption rate constants under convection-evaporation-enhanced mechanism (Section 5). The main conclusions are drawn in Section 6. The obtained results have several potential applications [34].

## 2. Materials and methods

*2.1. Materials*

In all our experiments below, we used three volatile amphiphiles (benzyl acetate, linalool, and citronellol) at fixed temperature of $T = 25\ °C$. The chemical structures of the amphiphiles are shown in Fig. 1. Benzyl acetate was a product of TCI (>99%, Cat. No. A0022): molecular mass $M_w = 150.18$ g/mol; density $\rho = 1054$ g/dm$^3$; solubility in water 3.1 g/dm$^3$, solubility limit $C_{sol} = 20.6$ mM; specific volume $1/v_m = \rho/M_w = 7.02$ M. To estimate the value of the diffusion coefficient of benzyl acetate in water, $D_d$, we calculate the equivalent spherical radius, $r_m$, from the specific volume, $v_m$: $r_m = 3.84$ Å. The dynamic viscosity of water at 25 $°C$ is $\eta_w = 0.889$ mPa·s, hence from the Stokes-Einstein law for diffusion in simple solutions, the calculated diffusion coefficient of the benzyl acetate in water is $D_d = 6.40 \times 10^{-10}$ m$^2$/s (see Table 1). The diffusion coefficient of benzyl acetate vapor in air, $D_v$, at 25 $°C$ is measured in Ref. [35]. In the literature the reported values of the saturation vapor pressure, $P_{sat}$, vary from 20 Pa to 190 Pa (at 25 $°C$), which means that the reliability of the available data for $P_{sat}$ can be questionable. For that reason in Appendix A we interpolated the experimental data for $P_{sat}$ at different temperatures and obtained that $P_{sat} = 21.8$ Pa at 25 $°C$, which is close to the result of 21.86 Pa reported in Ref. [36]. Hence the saturation concentration in vapor at 25 $°C$ is low, $C_{sat} = 8.79$ μM (see Table 1).

Linalool was a product of Sigma Aldrich (>97%, Cat. No. L2602) with molecular mass 154.25 g/mol, density and solubility in water at 25 $°C$ – 863 g/dm$^3$ and 1.589 g/dm$^3$, respectively. The values of $v_m$, $r_m$, and $D_d$ are listed in Table 1. The diffusion coefficient of linalool in air is measured in Ref. [37] and the saturation pressure is reported in Ref. [36]. One sees that linalool and benzyl acetate have similar physicochemical parameters (Table 1) but quite different chemical structures (Fig. 1) – as a result their surface activities are very different (see Section 3).

Citronellol was a product of Sigma (>95%, Cat. No. W230901): molecular mass 156.27 g/mol, density 855 g/dm$^3$; solubility in water 0.307 g/dm$^3$. Other physicochemical parameters are given in Table 1. We assume that the diffusion coefficient, $D_v$, of citronellol is close to those for other two amphiphiles. The experimental data for $P_{sat}$ at different temperatures are interpolated in Appendix A to estimate $P_{sat}$ at 25 $°C$. Citronellol has the lowest solubility in water and the lowest saturation concentration in vapor but the highest surface activity (Section 3).

The aqueous solutions were prepared with deionized water of specific resistivity 18.2 MΩ·cm purified by Elix 3 water purification system (Millipore). All experiments were carried out at a temperature of 25 °C.

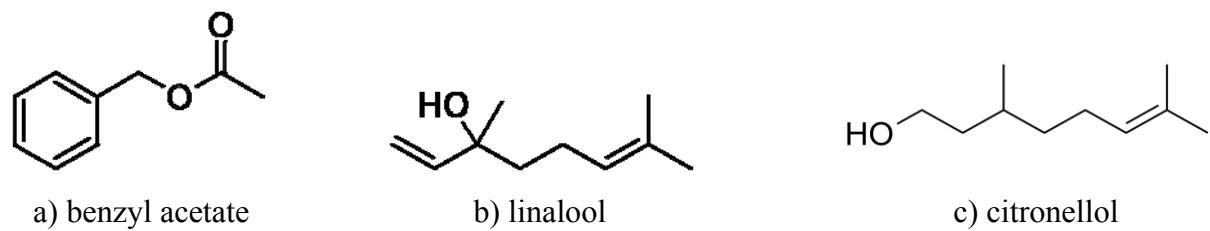

a) benzyl acetate      b) linalool      c) citronellol

**Fig. 1**. Chemical structure of volatile amphiphiles.

**Table 1**. Physicochemical properties of volatile amphiphiles at 25 °C.

|  | benzyl acetate | linalool | citronellol |
|---|---|---|---|
| $M_w$ (g/mol) | 150.18 | 154.25 | 156.27 |
| $\rho$ (kg/m³) | 1054 | 863 | 855 |
| $C_{sol}$ (mM) | 20.6 | 10.3 | 1.96 |
| $1/v_m$ (M) | 7.02 | 5.59 | 5.47 |
| $r_m$ (Å) | 3.84 | 4.14 | 4.17 |
| $\eta_w$ (Pa·s) | $8.89\times10^{-4}$ | $8.89\times10^{-4}$ | $8.89\times10^{-4}$ |
| $D_d$ (m²/s) | $6.40\times10^{-10}$ | $5.94\times10^{-10}$ | $5.89\times10^{-10}$ |
| $D_v$ (m²/s) | $6.00\times10^{-6}$ | $5.80\times10^{-6}$ | $\approx 5.9\times10^{-6}$ |
| $P_{sat}$ (Pa) | 21.8 | 22.1 | 7.15 |
| $C_{sat}$ (μM) | 8.79 | 8.92 | 2.88 |

*2.2. Experimental methods and protocols*

*Surface tension isotherms*. The stock aqueous solutions of volatile amphiphiles were prepared at concentrations equal to the solubility limit, $C_{sol}$. The stock solution was diluted to the desired concentration $C < C_{sol}$ and kept in a closed vessel in a thermostat at 25 °C for 24 hours. The surface tension was measured using the maximum bubble pressure method on BP 2 automated bubble pressure tensiometer (Krüss GmbH, Germany). To obtain the equilibrium surface tension, $\sigma_{eq}$, we used the long time asymptotic expansion equation [38]:

$$\sigma(t_{age}) = \frac{b_\sigma + \sigma_{eq} t_{age}^{1/2}}{a_\sigma + t_{age}^{1/2}} \qquad (1)$$

where $\sigma(t_{age})$ is the surface tension measured at the nominal surface age, $t_{age}$. The meaning of $t_{age}$ is the time shown by the apparatus (not subjected to corrections), $a_\sigma$ and $b_\sigma$ are constants. Note that Eq. (1) is valid for a diffusion controlled mechanism of adsorption. The obtained experimental surface tension isotherms, $\sigma_{eq}(C)$, are shown in Section 3.

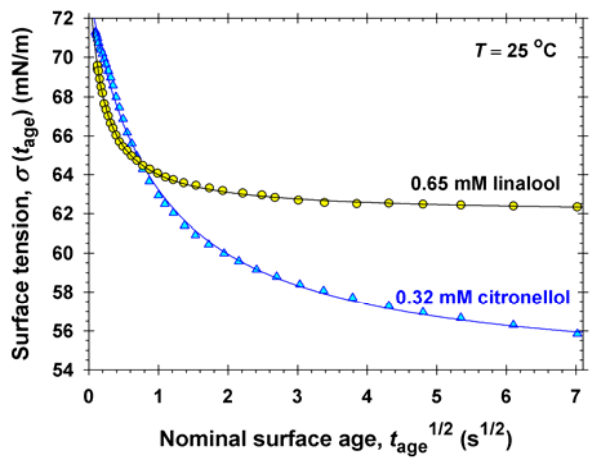

**Fig. 2**. Surface tension vs nominal surface age for 0.65 mM linalool (○) and 0.32 mM citronellol (Δ) measured using the MBPM. The solid lines show the best fit using Eq. (1), from which the equilibrium surface tension, $\sigma_{eq}$, is calculated.

Fig. 2 shows typical experimental data for $\sigma(t_{age})$ – the solid lines therein represent the best fit using Eq. (1). One sees that: i) the mechanism of adsorption corresponds to diffusion control; ii) the two times lower concentration of citronellol (0.32 mM) leads to the lower value of $\sigma_{eq}$, but the surface tension relaxation is considerably slower than that for linalool (0.65 mM). In all cases the values of the regression coefficients were greater than 0.9995 and the precision of the calculated equilibrium surface tension, $\sigma_{eq}$, was 0.1 mN/m.

*Adsorption from vapor – evaporation from drop.* The dynamics of adsorption-desorption of the surface active volatile amphiphiles were studied using the pendant drop method. All measurements were performed on DSA 100 R (Krüss GmbH, Germany) apparatus. The software DSA 1 fitted the experimental pendant drop profile with the Laplace equation of capillarity and calculated the surface tension, drop volume and area, and also the fit error. In all cases the fit error was small, which indicates that the vapor/solution interface is fluid [39]. Formation of adsorption multilayers at long times (up to 4000 s), like those for alkane [21,22], were not observed.

The experimental protocol was the following. In the first regime, which is the adsorption from vapor, the volatile amphiphile (benzyl acetate, linalool, and citronellol) was placed at the bottom of a small cuvette that was capped with a piece of filter paper socked with the given volatile species. The cuvette is placed in a temperature control chamber TC 40 (Krüss GmbH, Germany) at fixed temperature of 25 °C. The drop of deionized water was formed at the tip of metal needle with diameter 1.833 mm in the saturated atmosphere with the volatile amphiphile vapor. The surface tension decreases with time $t$ because of the adsorption of species from the vapor to the drop interface (Fig. 3). We waited different time (depending on the characteristic adsorption time of the given volatile amphiphile) for equilibration of the interfacial layer, manifested as reaching a steady-state value of the surface tension. In all cases, the drop volume, $V(t)$, was approximately constant (see Fig. 3b). In the first regime, we studied the *adsorption from vapor* to the liquid interface (see Section 4).

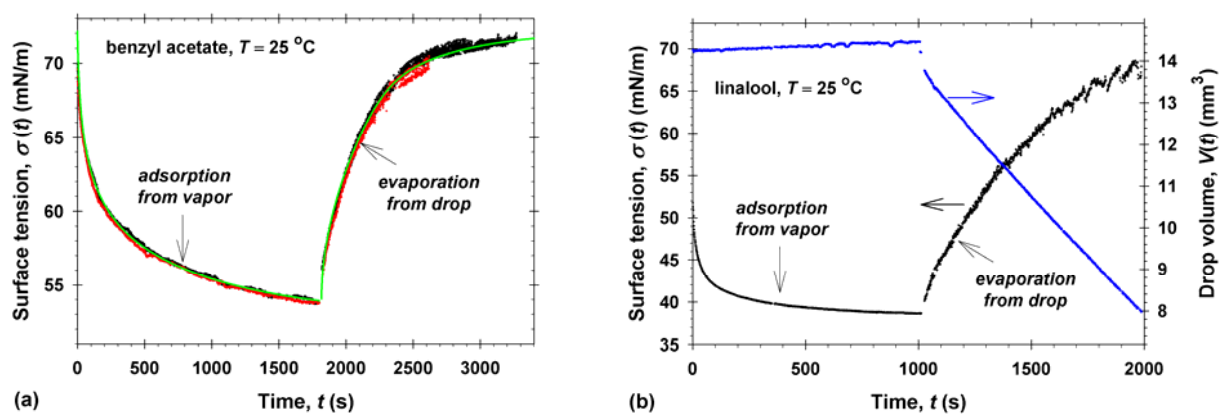

**Fig. 3**. Adsorption from vapor – evaporation from drop experiments (pendant drop method): a) illustration of the reproducibility, showing two independent experiments with benzyl acetate, represented by red and black dots; b) run with linalool – the black dots show $\sigma(t)$. The drop volume, $V(t)$, is displayed in blue; it stays approximately constant during the adsorption from vapor regime. One can see that $V(t)$ decreases with time in the regime of evaporation from drop. The green solid line in Fig. 3a shows the best theoretical fit, see Sections 4 and 5.

Subsequently, in the second regime (*evaporation from drop*), we oppened the temperature control chamber and removed the cuvette as fast as it was possible (for about 10 s). The same drop became in contact with the ambient atmosphere in the room (a large resevoir without vapour of volatile amphiphiles) at 25 °C. Because of the evaporation, the drop volume decreases with time (see Fig. 3b) and the measured surface tension, $\sigma(t)$, increases up to the surface tension of pure water 72 mN/m. The increase of $\sigma(t)$ is an indirect measure for the kinetics of the volatile amphiphile evaporation (see Section 5).

The experiments with benzyl acetate and citronellol were repeated at least three times and those with linalool – six times. The reproducibility of the kinetics curves in both stages (adsorption from vapor and evaporation from drop) was excellent. Fig. 3a shows the typical reproducibility between different experiments with volatile amphiphiles (for example in the case of benzyl acetate). Fig. 3b illustrates the typical drop volumes and the rate of drop evaporation (the slope of the drop volume versus time) after putting the drop in contact with the atmosphere without saturated vapor in its surroundings (at time $t > 1000$ s). The significant difference between the kinetics of adsorption of benzyl acetate and linalool is well illustrated: i) the process of adsorption from vapor of benzyl acetate is much slower than that of linalool, compare the left branches of the $\sigma(t)$ curves in Figs. 3a and 3b; ii) for benzyl acetate, the change of $\sigma(t)$ in the regime of evaporation from drop is faster than that in the regime of adsorption from vapor (Fig. 3a), while for linalool this trend is exactly opposite (Fig. 3b). The detailed theoretical and experimental explanations of these phenomena are given in Sections 4 and 5.

## 3. Surface tension isotherms of volatile amphiphile aqueous solutions

The experimental surface tension isotherms of volatile amphiphile aqueous solutions are processed using the van der Waals type of adsorption model [40,41]:

$$KC = \alpha \Gamma f(\Gamma) \text{ and } f(\Gamma) \equiv \frac{1}{1-\Gamma\alpha} \exp(\frac{\Gamma\alpha}{1-\Gamma\alpha} - \frac{2\beta}{k_B T}\Gamma) \tag{2}$$

Here: $k_B$ is the Boltzmann constant; $\Gamma$ is the adsorption; $K$ is the equilibrium adsorption constant related to the aqueous phase; $\alpha$ is the minimal (or "excluded") area per molecule; $\beta$ is the interaction parameter, which is positive for attraction between the adsorbed molecules in lateral direction; $f(\Gamma)$ is the surface activity coefficient, which accounts for the hard core and long range molecular interactions of adsorbed molecules. The expression for the two-dimensional equation of state corresponding to the van der Waals model reads:

$$\sigma = \sigma_0 - k_B T \frac{\Gamma}{1-\Gamma\alpha} + \beta \Gamma^2 \tag{3}$$

where $\sigma_0 = 72.2$ mN/m is the surface tension of pure water at the given temperature. The adsorption constant is directly related to the free energy of adsorption, $E$, by the following relationship:

$$K = v_m \exp(\frac{E}{k_B T}) \tag{4}$$

where $v_m$ denotes the molar volume of the adsorbed species. In Ref. [18] the surface tension isotherms for linalool and citronellol aqueous solutions had been measured at 20 °C and processed using a long list of theoretical models. From our viewpoint, the most adequate results are obtained applying the van der Waals model, which is self-consistent with the kinetic curves reported in Sections 4 and 5.

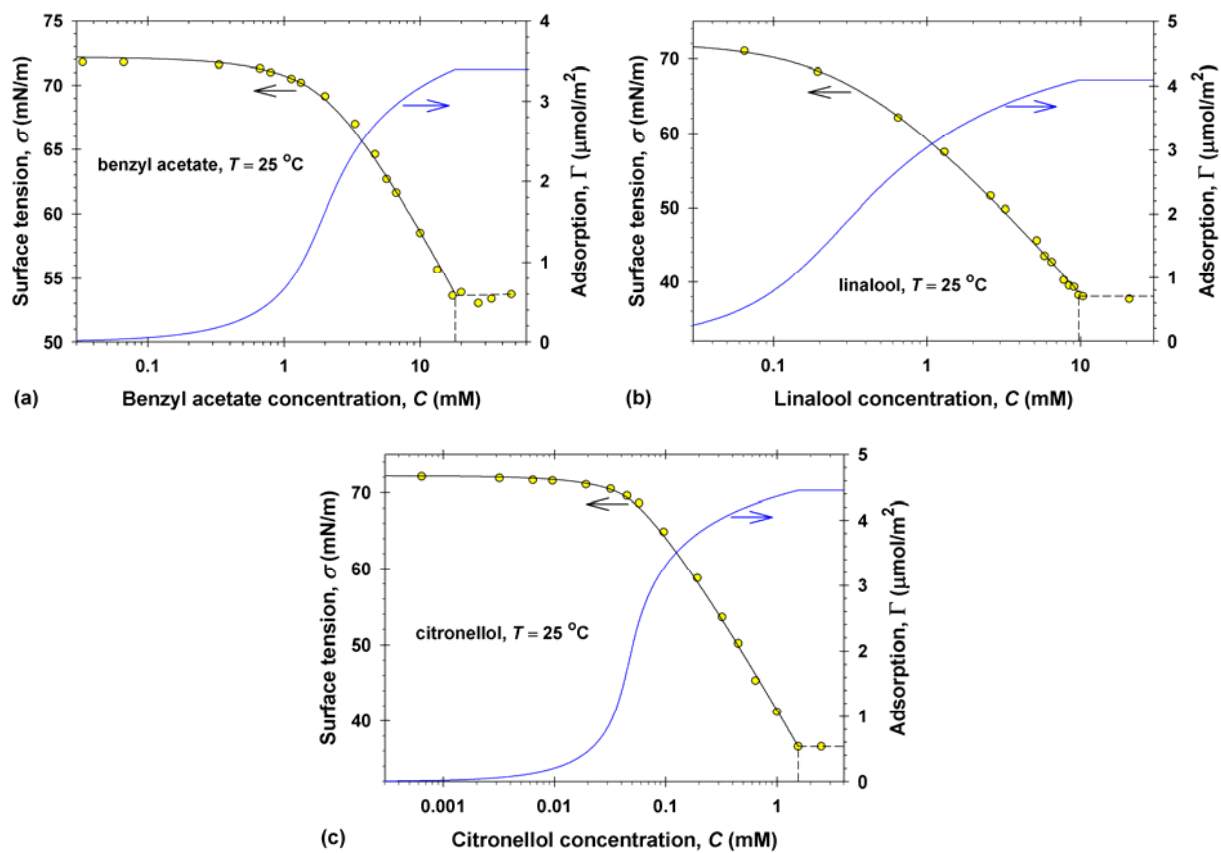

**Fig. 4**. Experimental (symbols) versus theoretical (lines) surface tension isotherms: a) benzyl acetate; b) linalool; c) citronellol. The right-hand side ordinate axes refer to the blue curves, representing the adsorbed amount per unit area, $\Gamma(C)$, as calculated implicitly from the combination of Eqs. (2) and (3).

Fig. 4 summarizes the experimental surface tension isotherms (symbols) for the three studied volatile amphiphiles. The solid lines therein show the best theoretical fits using the adjustable parameters, $K$, $\alpha$, and $\beta$, in Eqs. (2) and (3). The obtained best fit parameters of the van der Waals isotherm are listed in Table 2. The values for $K$ and $\alpha$ in the case of linalool and citronellol practically coincide with those reported in Ref. [18] for 20 $^{\circ}$C. Our values of the interaction parameter, $\beta$, are slightly different.

Constant values of the surface tension versus $C$ are measured for concentrations of benzyl acetate, linalool, and citronellol larger than 18 mM, 9.8 mM, and 1.6 mM, respectively (Fig. 4). These concentrations are close to the literature data for the solubility limits, $C_{sol}$, of the respective volatile amphiphiles (see Table 1). Therefore, the absence of change of $\sigma$ versus $C$ at high $C$ can be explained with constant chemical potential caused by aggregation in the bulk.

The solid lines in blue (reckoned at the right axes) in Fig. 4 show the predicted values of the adsorption, $\Gamma$, versus concentration $C$.

**Table 2**. Theoretical parameters of volatile amphiphiles at 25 $^{\circ}$C, calculated from the surface tension isotherms, and from the two kinetic regimes – adsorption from vapor and evaporation from drop.

|  | benzyl acetate | Linalool | citronellol |
|---|---|---|---|
| $\alpha$ (Å$^2$) | 35.6 | 30.5 | 30.2 |
| $E$ ($k_BT$) | 6.64 | 9.05 | 9.80 |
| $K$ (M$^{-1}$) | 109.0 | 1524 | 3297 |
| $\beta/(\alpha k_BT)$ | 2.05 | 0.965 | 2.52 |
| $k_{v,ads}$ (mm/s) | 2.90 | 11.8 | 0.785 |
| $k_{v,des}$ (s$^{-1}$) | 2.43 | 1.22 | 0.0637 |
| $k_{ev}$ (s$^{-1}$) | 5.90 | 0.358 | 0.105 |

The following conclusions can be drawn from the parameters of the van der Waals isotherms. The minimal excluded area per molecule, $\alpha$, has equal values for linalool and citronellol (30 Å$^2$), while for benzyl acetate – it is larger (36 Å$^2$). This result is in agreement with the chemical structure of the volatile amphiphiles (Fig. 1). In Ref. [41] the excluded area per molecule of linear alkylbenzene sulfonate (LAS) is obtained to be 37 Å$^2$. Hence, the latter value of $\alpha$ corresponds to the area of the benzene ring. The energy of adsorption, $E$, of benzyl acetate is about 2.5 $k_BT$ lower than that for linalool, and 3.2 $k_BT$ lower than that for citronellol. Hence, from the viewpoint of water solubility and surface activity, benzyl acetate is the most soluble and the least surface active volatile amphiphile, while citronellol is the least soluble and the most surface active species.

The obtained physicochemical parameters of the adsorption isotherms, listed in Table 2, are used in Sections 4 and 5 to calculate the surface tension theoretically from the adsorption, in order to explain the kinetic data for the surface tension in the regimes of adsorption from vapor and evaporation from drop.

## 4. Adsorption of volatile amphiphiles from vapor

For simplification of the numerical calculations in the case of kinetics of adsorption, we assume a spherical symmetry of the considered diffusion problem. Because of the spherical symmetry, the concentrations depend on time $t$ and radial coordinate $r$. A spherical drop with radius $a(t)$ contains an aqueous solution of volatile amphiphile with local bulk concentration $c_d(t,r)$ and diffusion coefficient $D_d$. The drop is immersed in vapors of the volatile amphiphile with local concentration $c_v(t,r)$ and diffusion coefficient $D_v$. The respective diffusion equations in the both environments are:

$$\frac{\partial c_d}{\partial t} = \frac{D_d}{r^2}\frac{\partial}{\partial r}(r^2 \frac{\partial c_d}{\partial r}) \text{ for } r < a(t) \text{ and } t > 0 \tag{5}$$

$$\frac{\partial c_v}{\partial t} = \frac{D_v}{r^2}\frac{\partial}{\partial r}(r^2 \frac{\partial c_v}{\partial r}) \text{ for } r > a(t) \text{ and } t > 0 \tag{6}$$

The general mass balance boundary condition at the drop interface relates the total fluxes of molecules from the drop, $j_d A$, and from vapor, $j_v A$, to the change of the total number of molecules at the interface, $\Gamma A$, with time:

$$\frac{d(\Gamma A)}{dt} = (j_d + j_v)A \text{ for } r = a(t) \text{ and } t > 0 \tag{7}$$

The concrete physicochemical meaning of the fluxes, $j_d$ and $j_v$, and the initial conditions are specified below.

In the case of *adsorption from vapor*, at the initial time, $t = 0$: i) the concentration in the vapor, $C_{sat}$, corresponds to the saturation vapor pressure, $P_{sat}$; ii) the aqueous phase does not contain the volatile amphiphile; iii) the adsorption process starts from a clean surface:

$$c_d(0,r) = 0 \text{ for } r < a \text{ ; } c_v(0,r) = C_{sat} \text{ for } r > a \text{ ; } \Gamma(0) = 0 \tag{8}$$

The drop volume and area do not change with time in this regime, so that the radius, $a(t) = a$, and the drop area, $A(t) = A$, are known constants from experiments (see Fig. 3b).

Note that the diffusion coefficient in the vapor phase is much higher than that in the aqueous solution, $D_v/D_d \approx 10^4$, but the concentration in the vapor phase is considerably lower, $C_{sat}/C_{sol} \approx 10^{-3}$ (see Table 1). Hence, it is difficult to compare preliminary the characteristic times of adsorption/desorption in both phases. For that reason, we chose the following strategy for theoretical modeling of the studied processes.

First, we assumed that the adsorption-desorption is so fast that the kinetics is purely diffusion controlled in both environments. Thus, the fluxes $j_d$ and $j_v$ are the respective diffusion fluxes:

$$j_d = -D_d \frac{\partial c_d}{\partial r} \text{ and } j_v = D_v \frac{\partial c_v}{\partial r} \text{ for } r = a \text{ and } t > 0 \tag{9}$$

In the diffusion controlled regime, contiguous subsurfaces (from the vapor and aqueous solution phases) are in chemical equilibrium with the interfacial phase and the respective chemical potentials are equal. Thus the subsurface concentrations, $c_{d,s}(t) = c_d(t,a)$ and $c_{v,s}(t) = c_v(t,a)$, are related to the partition coefficient:

$$\frac{c_{d,s}(t)}{c_{v,s}(t)} = \frac{C_{sol}}{C_{sat}} \quad \text{for } t > 0 \tag{10}$$

and the adsorption isotherm, Eq. (2), is valid with $C = c_{d,s}(t)$. In this case the diffusion problem, Eqs. (5)–(10), has no adjustable parameters and one can solve it numerically to obtain the kinetics of adsorption curve $\Gamma(t)$. From $\Gamma(t)$, one predicts $\sigma(t)$ using Eq. (3) and compares the obtained results with experimental data.

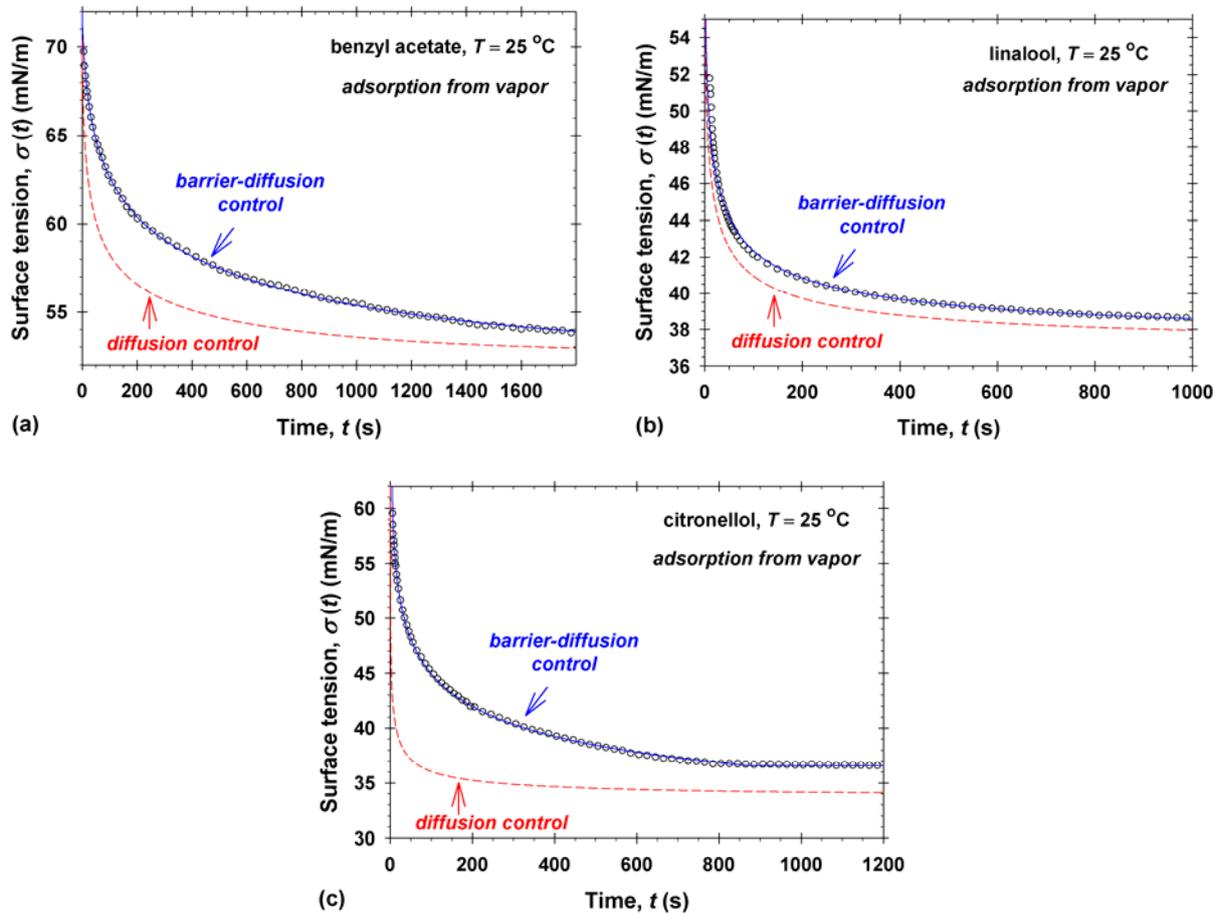

**Fig. 5**. Relaxation of the surface tension in the course of adsorption from vapor regime: a) benzyl acetate; b) linalool; c) citronellol. The symbols are experimental data, the dashed lines correspond to the diffusion controlled adsorption model, the solid lines show the best theoretical fit using the mixed barrier-diffusion controlled adsorption from the vapor and diffusion control from the drop phase.

We solve numerically the diffusion problem, Eqs. (5)–(10), using the Crank-Nicolson method keeping the nonlinearity of the boundary condition for the adsorption, Eqs. (2) and (7), at each time step [42]. This method is of the second order with respect to the time step,

Δ*t*. The radial coordinate is divided in a regular mesh with step Δ*r* = *a*/200 thus having 200 points in the drop and 4000 points in the vapor – the vapor phase is assumed bounded with large enough radius 20*a*. All space derivatives are interpolated up to the second order precision with respect to Δ*r*. The time step is Δ*t* = $10^{-5}$ s because of the large values of $D_v$.

Fig. 5 shows experimental data (symbols) for the relaxation of the surface tension, $\sigma(t)$, in the adsorption from vapor regime. For the sake of better illustration, not all experimental points like those shown in Fig. 3 are plotted. The dashed lines therein (Fig. 5) correspond to the numerical solution of the diffusion problem without adjustable parameters. One sees that for benzyl acetate and citronellol, the theory predicts considerably faster relaxation of the surface tension, while for linalool the theoretical line is closer to the experimental data. If one assumes that the desorption of the volatile amphiphile from the surface to the aqueous phase is slow, then the predicted surface tension relaxation becomes even faster. Hence, in order to describe the experimental data, one should account for the possible barrier adsorption mechanism in the vapor phase.

In the literature [21,22,30,31] the authors showed that the dynamics of adsorption of alkane vapor at the drop interface follows the barrier mechanism instead of the diffusion control. If one assumes that the adsorption of volatile amphiphile molecules from the vapor phase to the surface is slower and/or comparable to the diffusion, then the contiguous vapor phase is not in equilibrium with the interfacial phase and the boundary condition, Eq. (10), is not fulfilled. The mass balance of fluxes in the vapor phase requires the vapor diffusion flux, $j_v$, to be equal to the difference between the adsorption flux from vapor to the surface, $j_{v,ads}$, and the desorption flux from surface to the vapor, $j_{v,des}$ [43]:

$$j_{v,ads} - j_{v,des} = j_v = D_v \frac{\partial c_v}{\partial r} \quad \text{for } r = a \text{ and } t > 0 \tag{11}$$

The concrete expressions for $j_{v,ads}$ and $j_{v,des}$ depend on the mechanism of adsorption assumed to describe the surface tension isotherm (localized, non-localized, etc.) [44,45]. In the case of the van der Waals type of isotherms (non-localized adsorption), one defines [44,45]:

$$j_{v,ads} - j_{v,des} = k_{v,ads}[c_{v,s} - \frac{\Gamma \alpha}{K_v} f(\Gamma)] \tag{12}$$

where $k_{v,ads}$ is the adsorption rate constant and $K_v$ is the equilibrium adsorption constant corresponding to the vapor phase. In fact Eq. (12) introduces one unknown parameter – the adsorption rate constant, $k_{v,ads}$. The equilibrium constant $K_v$ is directly related to the solubility limit, saturation concentration, and equilibrium constant *K*, already obtained from the fit of isotherms (Table 2). The desorption rate constant, $k_{v,des}$, is defined from Eq. (12) as a coefficient of proportionality. So that:

$$K_v = K \frac{C_{sol}}{C_{sat}} \quad \text{and} \quad k_{v,des} = \alpha \frac{k_{v,ads}}{K_v} \tag{13}$$

The mixed barrier-diffusion control model in the vapor phase leads to the numerical solution of Eqs. (5) and (6) with initial conditions given by Eq. (8). The boundary conditions for the problem are: i) the mass balance equation (7) with the definitions of the diffusion fluxes, Eq. (9); ii) the adsorption isotherm, Eq. (2), in which $C = c_{d,s}(t)$; iii) the mass balance of fluxes in the vapor phase, Eq. (11), with the concrete form of the adsorption/desorption fluxes, Eq. (12).

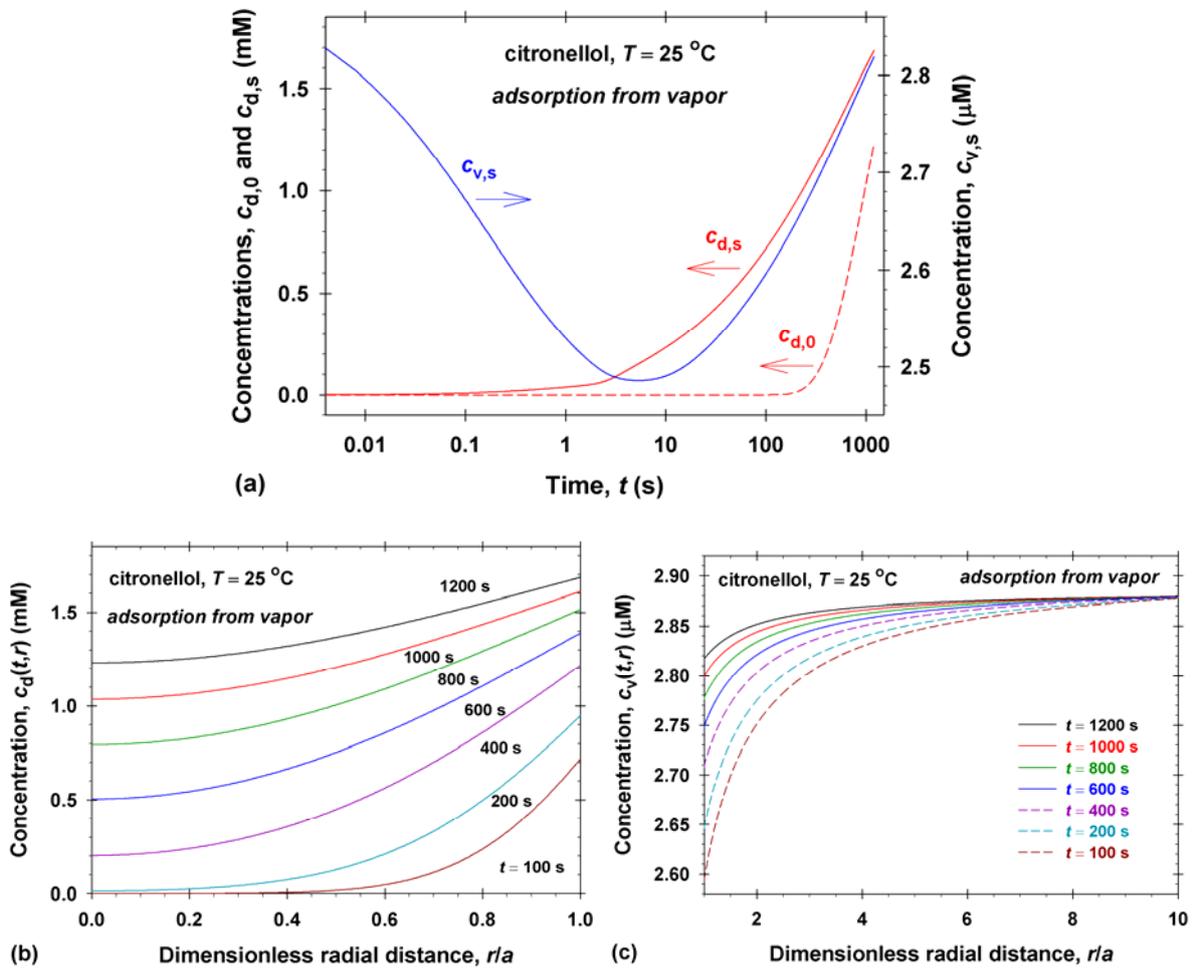

**Fig. 6**. Results from calculations for citronellol concentrations in the course of adsorption from vapor regime: a) dependencies of $c_{d,0}$, $c_{d,s}$, and $c_{v,s}$ on time; b) profiles of concentration in the drop phase, $c_d(t,r)$; c) profiles of concentration in the vapor phase, $c_v(t,r)$.

The solid lines in Fig. 5 correspond to the best theoretical fit applying the mixed barrier-diffusion control. The description of experimental data is excellent using one adjustable parameter $k_{v,ads}$. The calculated values of $k_{v,ads}$ and $k_{v,des}$ for the three studied volatile amphiphiles are summarized in Table 2. The following general conclusions can be drawn. First, the linalool molecules have the fastest adsorption from the vapor to the surface, those of benzyl acetate are 4 times slower, and the smallest is the adsorption rate constant for citronellol – 15 times smaller compared to that of linalool. This result is well illustrated in Fig. 5. The desorption time of molecules from surface to the vapor is characterized by the value of the inverse desorption constant, $1/k_{v,des}$. From the viewpoint of the characteristic desorption time, the citronellol, linalool, and benzyl acetate molecules escape the surface for 15.7 s, 0.820 s, and 0.412 s, respectively (Table 2). This order of the volatile amphiphiles corresponds to their surface activity and surface tension isotherms (Fig. 4).

Fig. 6 summarizes the results from calculations for citronellol concentrations in the course of adsorption from vapor regime. The subsurface concentration in the vapor phase, $c_{v,s}(t)$, initially decreases and has a minimum for $t = 5.32$ s (Fig. 6a). For $t > 5.32$ s, the diffusion from vapor becomes operative and $c_{v,s}$ increases. Note that $c_{v,s} < C_{sat}$ and the distribution of concentration in the vapor phase, $c_v(t,r)$, is not uniform (Fig. 6c). The subsurface concentration in the drop phase, $c_{d,s}(t)$, and that in the center of drop, $c_{d,0}(t) = c_d(t,0)$, gradually increase over time (Fig. 6a). Even at $t = 1200$ s, the profile of concentration in the drop phase, $c_d(t,r)$, is not uniform and the system becomes closer and closer to the equilibrium state with time (Fig. 6b). The respective physical pictures for benzyl acetate and linalool are quite similar.

The obtained values of $k_{v,ads}$ for alkanes [22] are $\approx 10^{-9}$ m/s, while our given in Table 2 are $\approx 10^{-3}$ m/s. One sees that there is six orders of magnitude difference between the values of $k_{v,ads}$ for alkane and for the three volatile amphiphiles studied here. Note that the vapor pressure of citronellol is 7.15 Pa and the saturation concentration in vapor is 2.88 μM (Table 1). In contrast, the saturation pressure of heptane is 6.13 kPa and the saturation concentration in vapor is 2.47 mM. For short times, the change of adsorption over time, $d\Gamma/dt$, is approximately equal to $k_{v,ads}C_{sat}$. In order to have the same values for $d\Gamma/dt$, the adsorption rate constant of heptane should be three orders of magnitude smaller than that of citronellol. From the physicochemical viewpoint, the different values of the equilibrium adsorption constants of citronellol, $K_v = 2244$ m³/mol, and that of heptane, $K_v = 1.21$ m³/mol (see Ref. [22]) show that the energy of adsorption of citronellol is considerably larger than that of heptane.

## 5. Evaporation of volatile amphiphiles from drop

In the second regime, the vapor phase is removed and the drop with dissolved volatile amphimphile becomes in contact with the ambient atmosphere in the room at 25 $^{o}$C. The water vapor pressure is different than the saturation one, and the drop shrinkages with time – the drop area, $A(t)$, and volume, $V(t)$, both decrease. The DSA 100 R apparatus recorded the geometrical drop parameters with time resolution of 0.1 s (see Fig. 3b). In the mass balance boundary condition, Eq. (7), the surface deformation, $d \ln A / dt$, and the change of the drop radius with time, $a(t)$, are accounted for. In all studied cases the drop radii were between 1.8 mm and 2.2 mm. We interpolated the experimental values of $A(t)$ and $a(t)$ and used the respective interpolations in numerical calculations. Different drops have different experimental geometrical parameters. Fig. 7a shows the changes of experimental drop areas with time for the fastest and the slowest evaporation drops (symbols). One sees that the used cubic polynomial interpolation (solid line) describe $A(t)$ with an excellent precision. Note that the changes of ln$A$ in all experimental cases are not considerably different and they lay between the plotted curves in Fig. 7a.

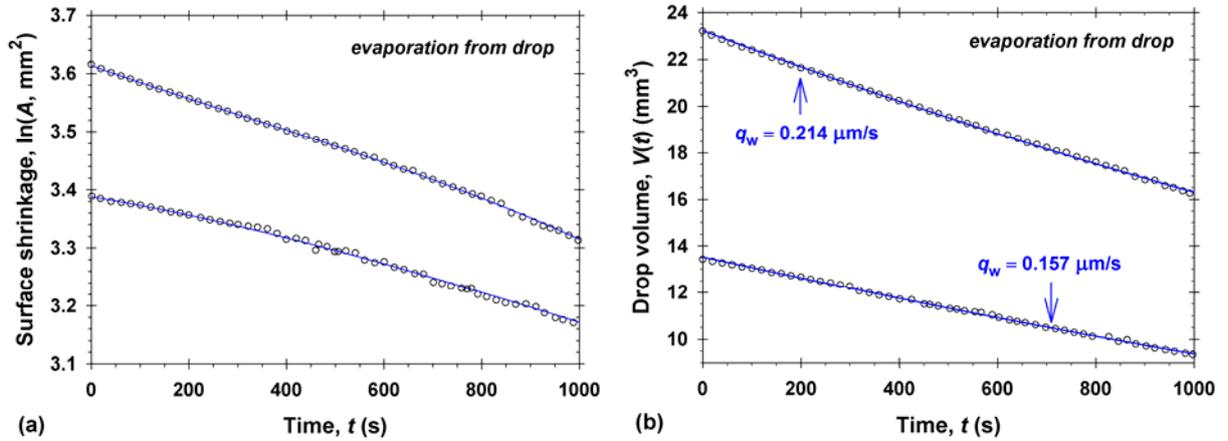

**Fig. 7**. Geometrical drop parameters versus time: a) decrease of surface area, $A(t)$, with time; b) change of drop volume, $V(t)$, with time.

In Section 4, the parameters of the barrier-diffusion control in the vapor phase are obtained (Table 2). In the open atmosphere, the air convection is considerably faster than the diffusion. Hence, in the *evaporation from drop* regime, we continue the calculations replacing the concentration, $c_v$, with zero and without adjustable parameters. Thus one solves numerically: i) the bulk diffusion equation in the drop phase, Eq. (5); ii) the initial condition for $c_d$ is taken to be the final distribution of the concentration in the drop, as calculated in the "adsorption from vapor" regime (see Fig. 6b); iii) the boundary condition is the mass balance

Eq. (7), in which flux $j_d$ corresponds to the diffusion flux, Eq. (9), and flux $j_v$ corresponds to the desorption flux, $j_{v,des}$, given by Eq. (12); in the later equation, the adsorption flux is $j_{v,ads} = 0$ because of $c_v = 0$. The calculated dependencies of the surface tension on time are plotted in Fig. 8 (dashed lines).

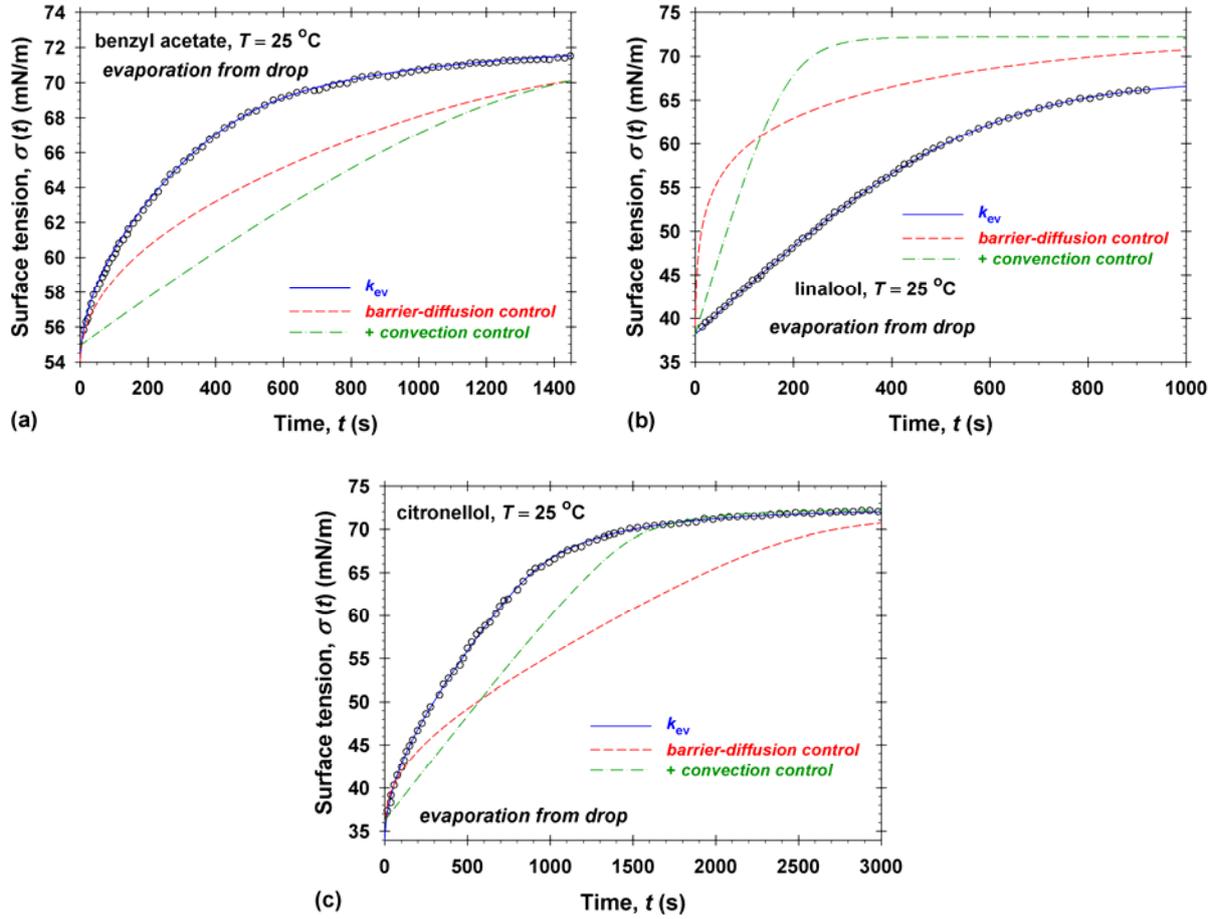

**Fig. 8**. Increase of the surface tension in the evaporation from drop regime: a) benzyl acetate; b) linalool; c) citronellol. The symbols are experimental data, the dashed lines correspond to the barrier control in vapor (with diffusion control in the drop), the dot-dashed lines − to the convection control in the drop phase, and the solid lines show the best theoretical fit using the model described in the main text.

The comparison between experimental data (symbols) and the theoretical curves (dashed lines) for the increase of the surface tension with time in the "evaporation from drop" regime shows that there is an additional physical mechanism, which is not included in the barrier-diffusion control model described in Section 4. If one integrates the general mass balance equation for the volatile amphiphile over the drop volume and adds the obtained result to the balance of the adsorbed species at the drop surface, then one arrives at the following general integrated mass balance equation:

$$\frac{d}{dt}(\int_A \Gamma \, dA + \int_V c_d \, dV) = \int_A j_v \, dA \tag{14}$$

see also Eq. (7). It is shown in the literature [46–48] that under the conditions of drop volume change because of evaporation, expansion, and/or shrinkage, an intensive hydrodynamic convection appears inside the drop. The characteristic time scale of the convection transport is much shorter than the characteristic diffusion time, which leads to equilibration of the concentration in the drop. Hence in the *convection control*, the concentration in the drop is approximately uniform and $c_d$ depends only on time, $t$. In this case, the initial condition for $c_d(t)$ is the mean volume value of the final distribution of $c_d(t,r)$, as calculated in the "adsorption from vapor" regime. In the literature [49], this mechanism is called *convection-enhanced adsorption*. For uniform adsorption, $\Gamma$, and flux, $j_v$, distributed along the surface, convection control in water, and barrier control in the vapor phase, Eq. (14) considerably simplifies:

$$\frac{d}{dt}(\Gamma A + c_d V) = -k_{v,des} \Gamma f(\Gamma) A \tag{15}$$

see Eq. (12).

The dependencies of the drop area, $A(t)$, and volume, $V(t)$, on time, $t$, are measured (see Fig. 7). The adsorption, $\Gamma(t)$, and volatile amphiphile concentration, $c_d(t)$, are related with the adsorption isotherm, Eq. (2), in which $C = c_d(t)$. The desorption rate constants, $k_{v,des}$, for the three volatile amphiphiles are given in Table 2. Thus one integrates numerically the differential equation, Eq. (15), without adjustable parameters and obtains the adsorption kinetics, $\Gamma(t)$. The relaxation of surface tension with time, $\sigma(t)$, is calculated from $\Gamma(t)$ and the two-dimensional equation of state, Eq. (3). The dot-dashed lines in Fig. 8 show the numerical results for the three studied amphiphiles. As can be expected, the convection in the drop phase decelerates the adsorption during the initial times, and accelerates the adsorption for the long times. Therefore, the convection-enhanced adsorption mechanism cannot explain experimental data for the surface tension relaxation in the evaporation from drop regime. The water evaporation plays an important role for desorption of the volatile amphiphiles from the surface.

The change of the drop volume with time is equal to the rate of water evaporation, $q_w$, multiplied by the drop surface:

$$\frac{dV}{dt} = -q_w A \quad \text{and} \quad \frac{d}{dt}(\frac{V}{v_w}) = -\frac{q_w}{v_w} A \tag{16}$$

where $v_w$ is the volume of the water molecule. The decrease of the drop area with time is interpolated in Fig. 6a. Subsequently, one fits experimental data for $V(t)$, symbols in Fig. 7b, with the numerical integration of Eq. (16) assuming a constant value of $q_w$. The solid lines therein (Fig. 7b) represent the best theoretic fit with $q_w$ = 0.157 μm/s for the slowest evaporating drop, and $q_w$ = 0.214 μm/s for the fastest evaporating drop. For all investigated drops, the volumes versus time curves lie between those plotted in Fig. 7b. Thus, the rate of water evaporation is a constant in the range from 0.157 to 0.214 μm/s for each individual drop.

In the literature, the Hertz-Knudsen equation is used to calculate the evaporation rate. The problems for theoretical prediction of $q_w$ are typically related to the temperature difference between liquid and vapor phases [50–53]. Hardly J.K. [54] reported the relationships between $q_w$ and the pressure differences for a constant temperature in the case of diffusion and convection limits. In our case, we measure $q_w$ and respectively, we know the right-hand side of Eq. (16). In the literature [50–54], the adsorption flux is typically called condensation and the desorption flux – evaporation. The physical meaning of the quantity $q_w A/v_w$ is the total number of water molecules, which leave the surface area $A$ per unit time due to the evaporation.

Fig. 8 shows that the calculated (dashed and dot-dashed lines) relaxations of the surface tension for benzyl acetate and citronellol are slower than the experimental trends. If one assumes a barrier mechanism of adsorption in the drop phases, then the predicted surface tension relaxation becomes even slower. The experimental results for linalool show the opposite trend. *Our hypothesis* is that the water flux from the surface transports also a certain part of the adsorbed volatile amphiphiles which have embarked on the drop surface; consequently, the desorption rate into the vapor phase will be affected. In the simplest case, the probability of $N_w$ water molecules at the surface to transport volatile amphiphile molecules increases with the number of the volatile amphiphile molecules at the surface, $N_s$. Hence, the expression for the total flux of the volatile amphiphile molecules from the surface because of the water evaporations, $J_{s,ev}$, can be written in the following form (corresponding to the van der Waals model):

$$J_{s,ev} = -(\frac{q_w}{v_w}A)\lambda \frac{N_s}{N_w} f(\Gamma) \tag{17}$$

where $\lambda$ is the coefficient of proportionality. This assumption is reasonable, because the more amphiphile molecules at the surface – the larger the flux is, $J_{s,ev}$. If there are no adsorbed molecules, then this flux is missing and $J_{s,ev} = 0$. One represents Eq. (17) in terms of the adsorption, $\Gamma$, as follows:

$$J_{s,ev} = -r_{ev}\Gamma f(\Gamma)A, \quad r_{ev} \equiv q_w \frac{\lambda \alpha_w}{v_w} \tag{18}$$

where $\alpha_w = A/N_w$ is the characteristic area of the water molecule at the surface and $r_{ev}$ is the respective rate constant measured in $s^{-1}$.

If we compare now the right-hand sides of Eqs. (15) and (18), then we will find that they have the same form – the difference is only on the multipliers, $k_{v,des}$ and $r_{ev}$, appearing therein. Hence, the water evaporation would affect the desorption rate constant from the drop surface to the vapor phase. For that reason, we repeated our calculations using Eq. (15), in which $k_{v,des}$ is replaced by the adjustable parameter, $k_{ev}$. The obtained best fit theoretical lines (solid lines) in Fig. 8 are drawn with the obtained values of $k_{ev}$ listed in Table 2. The agreement between the proposed theoretical model and experimental data is excellent for the three studied volatile amphiphiles. The ratios between the rate constants affected by water evaporation, $k_{ev}$, and the rate constants without water evaporation from the drop, $k_{v,des}$, are $k_{ev}/k_{v,des} = 2.43$, 0.293, and 1.65 for benzyl acetate, linalool, and citronellol, respectively. The water evaporation enhances the desorption of benzyl acetate and citronellol molecules from the drop surface to the vapor phase, while the same process of water evaporation decelerates the desorption of linalool.

The theoretical description of the experimental data for $\sigma(t)$ in both regimes, adsorption from vapor and evaporation from drop, as subsequent stages, is summarized in Fig. 3a (green solid line), for the case of benzyl acetate. The two adjustable parameters are $k_{v,ads}$ in the "adsorption from vapor" stage, and $k_{ev}$ in the "evaporation from drop" regime (Table 2). For the other two volatile amphiphiles, the respective kinetic curves look as successions of those depicted in Figs. 5b and 8b for linalool, and those in Figs. 5c and 8c for citronellol. The three values of the adsorption rate constants from vapor, $k_{v,ads}$, presented in Table 2, are within the order of $10^{-3}$ m/s, while the reported values for hexane are $\approx 10^{-9}$ m/s [54] and for cyclohexane they are $\approx 10^{-10}$ m/s [54].

## 6. Conclusions

Here, we investigated experimentally and theoretically the mass transfer between vapor and aqueous solution for three volatile amphiphiles (benzyl acetate, linalool, citronellol), which have limited (non-negligible) solubility in water. The necessary reliable physicochemical data to be used for the theoretical description, like diffusion coefficients in the vapor and aqueous phases, saturation vapor pressure and concentration, solubility limit in water, etc., are summarized in Table 1 in order to minimize the number of possible adjustable parameters. The experimental equilibrium surface tension isotherms are processed using the model of non-localized adsorption (the van der Waals type of isotherm). The relevant parameters in this isotherm – the energy of adsorption, $E$, the minimal area per molecule, $\alpha$, and the attraction energy of interaction between adsorbed molecules in lateral direction, $\beta$, are calculated, see Table 2 and Section 3. The excluded areas per molecule are approximately equal for linalool and citronellol (30 Å$^2$), while the benzene ring of benzyl acetate leads to the larger values of $\alpha = 36$ Å$^2$. The lowest adsorption energy of benzyl acetate corresponds to its highest solubility in water and the weakest surface active. On the other hand, for citronellol the surface activity and the adsorption energy are highest. The measured constant values of the surface tension versus amphiphile concentration above the solubility limit are result of the fixed chemical potential caused by aggregation in the bulk.

The dynamics of the volatile amphiphile adsorption from saturated vapor to a confine aqueous volume (drop) is measured and characterized by the relaxation of the surface tension. The accurate numerical calculations show that the diffusion control adsorption from both phases leads in all cases to faster relaxation than the experimentally observed. Taking into accounting a barrier mechanism in the vapor phase, simultaneously with diffusion (that is, barrier-diffusion control), leads to excellent theoretical description of the experiments (Fig. 5) with one adjustable parameter – the "adsorption from vapor phase" rate constant, $k_{v,ads}$. From the obtained equilibrium adsorption constant $K$ and from $k_{v,ads}$, the rate constant of desorption from the surface toward the vapor, $k_{v,des}$, is determined (Table 2). These physicochemical parameters completely characterize the adsorption/desorption of the studied three volatile amphiphiles. Because of the different chemical structures of benzyl acetate, linalool, and citronellol compared to alkanes, our values of $k_{v,ads}$ are $\approx 10^{-3}$ m/s, while those for hexane and cyclohexane are $\approx 10^{-9}$ m/s and $\approx 10^{-10}$ m/s, respectively [54]. The characteristic desorption times from the A/W adsorption layers to the vapor, $1/k_{v,des}$, are equal to 15.7 s, 0.820 s, and 0.412 s, respectively for citronellol, linalool, and benzyl acetate molecules, which explicitly correlates with their surface activity.

In a second scenario, aqueous drops containing benzyl acetate, linalool, or citronellol, which have been previously dissolved from the respective vapors, are placed in contact with clean air (without vapors of amphiphiles). Then, the mass transfer goes in the opposite direction, from the aqueous solution to the air. The volatile amphiphile's desorption is accompanied with the evaporation of water from the drop. From the increase of the surface tension, $\sigma$, with time, $t$, up to the values corresponding to pure water (Fig. 8), we draw conclusions for the effect of water evaporation on the desorption rate constant of the amphiphiles. The numerical calculations without adjustable parameters in the case of diffusion control or convection-enhanced control in the drop phase (plus $k_{v,des}$ in the vapor) showed that both mechanisms lead to slower surface tension relaxation for benzyl acetate and citronellol and to faster relaxation for linalool compared to experimental data. The experimental data for $\sigma(t)$ in the "evaporation from drop" regime are described excellently with convection-enhanced mechanism of adsorption from water, combined with barrier desorption from drop to vapor, whose rate constant, $k_{ev}$, is affected by the water evaporation. The latter quantity, $k_{ev}$, is left to be an adjustable parameter, and is determined from data fits. We obtain that $k_{ev}/k_{v,des}$ is equal to 2.43 for benzyl acetate and to 1.65 for citronellol, which explains their faster surface tension relaxation. Oppositely, for linalool $k_{ev}/k_{v,des}$ is equal to 0.293, and the water evaporation suppresses the evaporation of linalool. Note, that the mass transfer processes in the case of evaporation are complex: i) the evaporation of water effectively increases the concentration of amphiphiles in drop; ii) the evaporation of amphiphiles decreases their concentration in the aqueous phase; iii) the water and amphiphile fluxes from the surface to the air are interrelated. It is impressive, that these complex effects can be quantify by means of one parameter only – the rate constant, $k_{ev}$. To obtain the dependence of $k_{ev}$ on the physicochemical parameters of the studied system, future experiments with different humidity of the vapor phase are needed. Such experiments will clarify the effect of enhanced or suppressed water evaporation on the values of $k_{ev}$.

**Acknowledgements**

The authors are grateful to the project # DO 02/4 – 12.06.2018 with the Bulgarian Science Fund (FNI-MON), for the financial support. K. Danov acknowledges the support from the Operational Programme "Science and Education for Smart Growth", Bulgaria, project No. BG05M2OP001-1.002-0023.

**Appendix A. Saturation pressure of benzyl acetate and citronellol**

The Clausius-Clapeyron equation:

$$\ln[\frac{P_{sat}(T_2)}{P_{sat}(T_1)}] = -\frac{\Delta H_{vap}}{R}(\frac{1}{T_2}-\frac{1}{T_1}) \tag{A1}$$

relates the saturation pressure, $P_{sat}$, and temperature, $T$, where $R$ is the specific gas constant and $\Delta H_{vap}$ is the specific latent heat of evaporation. In Fig. A1 we plotted the experimental data for benzyl acetate [55,56] and citronellol [57] in accordance with Eq. (A1). From the linear regression fit, one obtains the equations inserted in Fig. A1. The predicted values of the saturation pressures for benzyl acetate and citronellol at 25 °C are shown by symbol □ and the obtained values are included in Table 1.

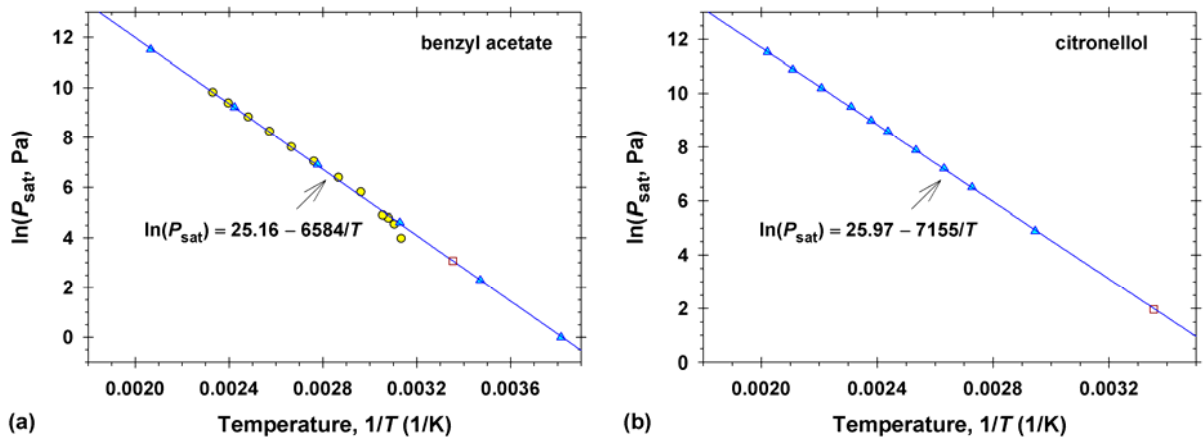

**Fig. A1**. Dependence of the saturation pressure on temperature: a) benzyl acetate, ○ data from Ref. [55], Δ data from Ref. [56], □ interpolation data for 25 °C (Table 1); b) citronellol, Δ data from Ref. [57], □ interpolation data for 25 °C (Table 1).